\documentclass[11pt]{llncs}
\usepackage{graphicx}
\usepackage{cite}
\usepackage{url}
\usepackage{times}
\usepackage{mathfont}

\def\max{\mathop{\rm max}}
\def\log{\mathop{{\rm log}}}
\def\inf{\mathop{{\rm inf}}}
\def\det{\mathop{{\rm det}}}

\def\vol{\mathop{{\rm vol}}}

\mathcode`O="724F

\DeclareSymbolFont{AMSb}{U}{msb}{m}{n}
\DeclareSymbolFontAlphabet{\Bbb}{AMSb}

\def\E{\ensuremath{\Bbb E}}

\DeclareMathSymbol{\subsetneq}{\mathrel}{AMSb}{"28}

\makeatletter
\def\hb@xt@{\hbox to }
\makeatother

\let\oldendproof\endproof
\def\endproof{\qed\oldendproof}

\begin{document}

\title{Optimized Color Gamuts for Tiled Displays}
\author{Marshall Bern\inst{1} and David Eppstein\inst{2}}
\authorrunning{Eppstein}
\institute{Palo Alto Research Ctr., 3333 Coyote Hill Rd., Palo Alto, CA 94304, USA,
\email{bern@parc.com}
\and
Information \& Computer Science, UC Irvine, CA 92697-3425, USA,
\email{eppstein@ics.uci.edu}}
\date{ }

\maketitle
\begin{abstract}
We consider the problem of finding a large color space that can be generated by all units in multi-projector tiled display systems.  Viewing the problem geometrically as one of finding a large parallelepiped within the intersection of multiple parallelepipeds, and using colorimetric principles to define a volume-based objective function for comparing feasible solutions, we develop an algorithm for finding the optimal gamut in time $O(n^3)$, where $n$ denotes the number of projectors in the system.  We also discuss
more efficient quasiconvex programming algorithms for alternative objective functions based on maximizing the quality of the color space extrema.
\end{abstract}

\section{Introduction}

Demand for increased image resolution has grown explosively in recent years, both in imaging devices such as cameras and in displays for those images.
In particular, collaborative workspaces require large display walls that can accomodate both presentation-sized graphics and higher resolution images for up-close viewing.
In order to satisfy this demand for display resolution, researchers and corporations have resorted to {\em tiled displays}, in which multiple display units (typically rear projection systems) each cover a small portion of the overall display area
\cite{HumHan-Vis-99,LiCheChe-CGA-00,RasBroYan-Vis-99}.
However, achieving a seamless appearance with such a system poses formidable technical challenges, involving careful geometric alignment of the individual display units, feathering or masking seams between units, correction for distortion, vignetting, and other defects of individual units' projection systems, and matching of the color characteristics among the different units in the system.

In this paper we address one of these challenges: finding an appropriate color gamut, achievable by all projectors in the system, so that scenes can be rendered on the system without distracting color shifts on the boundaries between different projectors' output.  This problem was previously considered by Stone~\cite{Sto-CGA-01}, who described a non-automated procedure for finding a large common color gamut for tiled displays.  Majumder et al.~\cite{MajHeTow-Vis-00} also studied color matching problems in tiled displays, however their paper addresses only the related problem of mapping a suitable device-independent gamut into individual projectors' color spaces, and does not discuss how to find such a gamut.  We show how to formalize this challenge as a computational geometry optimization problem, and describe efficient algorithms for solving various formulations of this problem.

\section{Color Gamuts}
\label{sec:color}

We review briefly the measurement and algebra of {\em additive color} devices such as LCD projectors; for more details see Stone~\cite{Sto-CGA-01} or her references.

In these devices, pixels are combinations of separate red, green, and blue light signals, each of which has its color determined by filtering light from a white light source, and its intensity determined by a single liquid crystal element.  When measured in a suitable device-independent color space, color values can be viewed as elements of a three-dimensional linear algebra: the superposition of two color signals has a value measured by the three-dimensional vector sum of the corresponding two color values.  Thus, the {\em gamut} of a three-signal additive color device (that is, the set of colors that it can display) naturally forms a parallelepiped in color space, in which the eight parallelepiped corners correspond to eight color values: black (which we denote $K$ for short), white ($W$), red ($R$), green ($G$), blue ($B$), cyan ($C$), magenta ($M$), and yellow($Y$).  However, it is not usually the case that the black point $K$ lies at the zero point of the vector space of colors, because a projector displaying black still projects some light onto its screen.  Given a suitable set of four of these vectors (say, $K$, $R$, $G$, and $B$), we can determine the other four extremal colors as linear combinations of these four: $W=K+(R-K)+(G-K)+(B-K)=R+G+B-2K$, $Y=K+(R-K)+(G-K)=R+G-K$, etc.

\begin{figure}[t]
\centering
\includegraphics{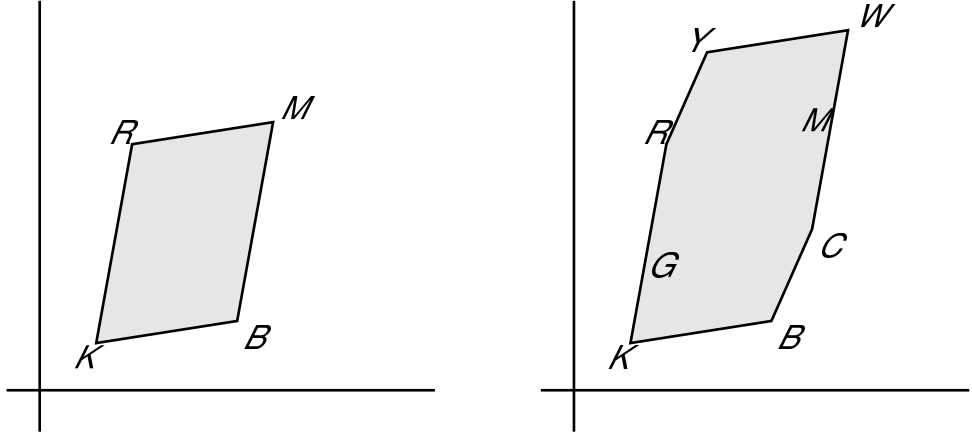}
\caption{Two-dimensional analogues of three-dimensional color gamuts.  The vertical and horizontal lines mark the coordinate axis of a two-channel device independent color space.  Left: a two-channel additive color device can produce a gamut of colors in the form of a parallelogram; analogously, a three-channel projector has a three-dimensional gamut in the form of a parallelopiped.  Right: a three-channel additive color device, when measured in a two-channel device independent color space, produces a gamut in the form of a hexagon; analogously, a four-channel projector in three-dimensional color space produces a rhombic dodecahedron gamut.}
\label{fig:2dgamuts}
\end{figure}

To provide a simplified view of this three-dimensional color space, our figures illustrate
two-dimensional analogues.  For instance, a person with red-green color blindness can only sense two of the three color bands seen with normal vision; color for such a person would naturally be measured in a two-dimensional device-independent color space, formed by projection from the usual three-dimensional color spaces, and indicated by the coordinate axes in the figures.  Similarly, a projector with a short circuit in the green color channel's signal would generate only a two-dimensional color gamut, in the form of a
parallelogram with four corners $K$, $R$, $B$, and $M=R+B-K$: see Figure~\ref{fig:2dgamuts}, left.

We note that some devices, including projector screens based on color wheels and mirror arrays instead of liquid crystals, may form their pixels from the superposition of four or more color signals; for instance certain mirror array displays have a fourth white position on their color wheels~\cite{Sto-CGA-01}.  If these signals are independently controllable, the resulting gamut has the form of a {\em zonotope} \cite{Epp-MER-96} such as a rhombic dodecahedron; to return to the two-dimensional simplified color model of our illustrations, a viewer with red-green color blindness viewing a three-color projector would see a gamut in the form of a hexagon in his two-dimensional color space (Figure~\ref{fig:2dgamuts}, right).
In practice, however, the white color is added in progressively as a function of the intensity of the input signal, resulting in gamuts in the form of elongated parallelepipeds.  Our algorithms can easily be adapted to handle such shapes without significantly increasing their complexity.

Due to variations in the manufacturing process for their filters, differences in the batch and age of the light bulbs used in their white light sources, and other factors, different projectors, even of the same model, will have gamuts that form different parallelepipeds in the device-independent color space.  If one uses these projectors in a tiled display without correcting for their different gamuts, the resulting difference in visual appearance will mar the appearance of seamlessness that most tiled display systems aim for.  Thus, we must find a {\em standard gamut} that contains only colors that can be displayed by all projectors in our system.

\begin{figure}[t]
\centering
\includegraphics{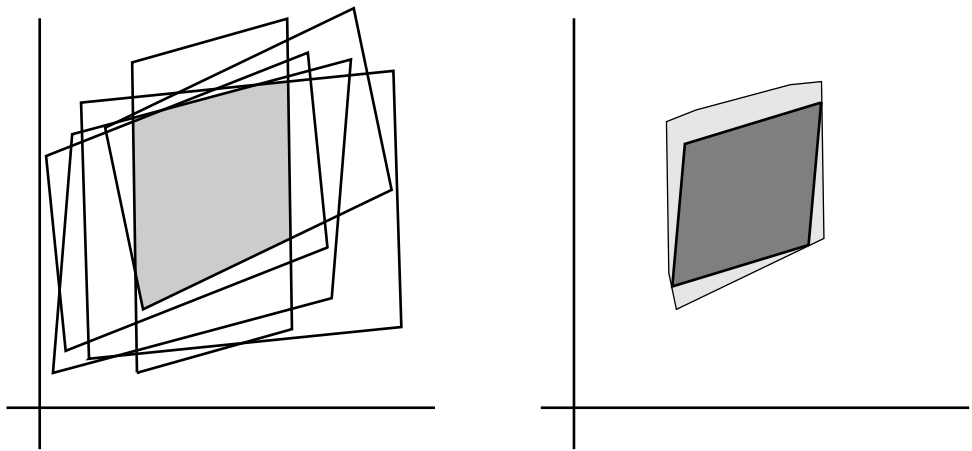}
\caption{Left: the intersection of projector gamuts forms an irregular convex shape that is not a parallelepiped (or a parallelogram in our two-dimensional simplified color model).  Right: the task is to find a large additive color gamut within the intersection of the projector gamuts.}
\label{fig:2dintersect}
\end{figure}

The most obvious approach to this standard gamut problem, from a computational geometry point of view, is to find the three-dimensional shape formed by intersecting the gamuts of all the individual projectors; if we have $n$ projectors, this shape can be represented as the intersection of $6n$ halfspaces (the faces of the parallelepipeds from each projector), and constructed in time $O(n\log n)$.  However, the resulting shape will likely be irregular and not itself be a parallelepiped (Figure~\ref{fig:2dintersect}, left).  Non-parallelepiped gamuts commonly arise in subtractive color devices such as inkjet printers, and there has been much work on mapping device-independent colors onto such gamuts~\cite{HarSch-CIC-97}, but there are significant technical difficulties in this approach:
Complicated algorithms are required to map colors from a device independent space (such as the sRGB~\cite{StoAndCha-W3C-96} typically used for computer graphics) into such gamuts, and the transformation from the device independent space to the gamut may introduce clipping or color shifts for saturated colors near the boundaries of the gamut.
There are difficulties even in using a relatively simple non-parallelepiped gamut such as a warped parallelepiped with curved, bilinear sides; for example, computer graphics algorithms such as anti-aliasing rely on the additivity of colors.

Instead, following Stone~\cite{Sto-CGA-01}, we wish to treat the tiled display system as if it were a single additive color device, by finding a large parallelepiped contained in the intersection of the projector gamuts (Figure~\ref{fig:2dintersect}, right), and using this parallelepiped as our standard gamut.
Each projector can represent all the colors in such a standard gamut, and the transformation
from the standard gamut in the device independent color space to the gamuts of the individual projectors can be performed in a particularly simple way, by a linear transformation represented by a $4\times 4$ matrix~\cite{Sto-CGA-01}.

Thus, we can state formally the problem of finding a standard gamut as a computational geometry optimization problem: given a suitable objective function $q$ (measuring the quality of a given gamut), and a set of $n$ parallelepipeds (projector gamuts), find a parallelepiped $P$ contained in the intersection of the projector gamuts and maximizing $q(P)$.
A natural choice for $q$ would be the volume of $P$, but this seems to lead to inefficient algorithms due to the high number of degrees of freedom, 12, in the specification of $P$.  Instead it seems appropriate to apply colorimetric principles to simplify the problem as well as to achieve a better match between the objective function and viewer-perceived quality. In the following sections, we explore the choice of objective function and its algorithmic implications.

\section{Stone's Procedure}

\begin{figure}[t]
\centering
\includegraphics{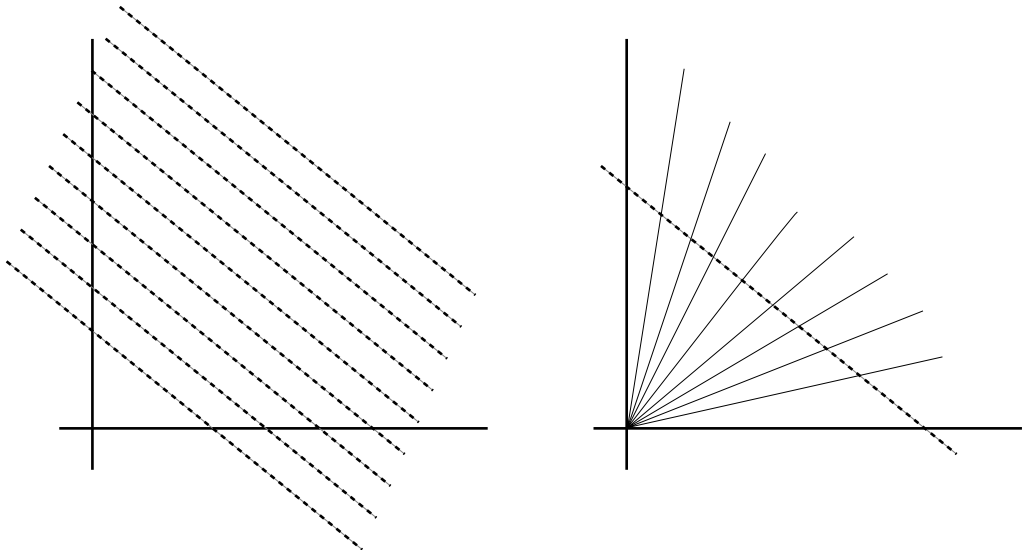}
\caption{Left: colors of fixed luminosities form parallel planes in color space.  Right: Lines of constant chromaticity radiate from the origin.}
\label{fig:chromaluma}
\end{figure}

Stone~\cite{Sto-CGA-01} described a hand-computation procedure for finding large standard gamuts.  Her procedure is based on factoring the three-dimensional device-independent color space into a two-dimensional chromaticity and one-dimensional luminosity.   The {\em luminosity} of a color value is measured by a positive linear combination of the values of its channels; sets of colors with given fixed luminosity values form parallel planes in the device independent color space (Figure~\ref{fig:chromaluma}, left).
From the parallelepiped representing any additive color device's gamut, we can form a {\em chromaticity diagram} by translating the black point to the origin of the color space and then projecting the translated gamut onto a plane of constant luminosity; the projection of the parallelepiped forms a triangle in chromaticity space.
The inverse image of any point in the projection plane is a line through the origin with constant chromaticity (Figure~\ref{fig:chromaluma}, right).

Stone's procedure consists of the following steps:

\begin{enumerate}
\item Intersect the chromaticity diagrams of the projectors, and find a large triangle within this intersection. Stone does not specify how this triangle is selected, but efficient algorithms are known for finding large triangles within convex polygons~\cite{AggPar-FOCS-88,BoyDobDry-SJC-85}.
\item Select the chromaticity of the black point $K$ of the standard gamut by averaging the chromaticities of the black points of the individual projectors.  Select $K$ itself as the darkest value of that chromaticity within the gamuts of all projectors.
\item Select the chromaticity of the white point $W$, to be the same as that of the black point and select $W$ as the brightest value of that chromaticity within the gamuts of all projectors.
\item Form a standard gamut by using the chromaticity diagram selected in step 1, the black point selected in step 2, and the white point selected in step 3.
\item The resulting gamut may have bright colors that fall outside the gamuts of the individual projectors.  If so, scale the gamut by reducing the luminosities of the primary colors until it is contained in the projector gamuts.
\end{enumerate}

Although all of these choices are well motivated, it is difficult to find a quality measure that they optimize, and some of the steps in this procedure would need to be more precisely specified in order to transform it into a computer algorithm.

\section{Constrained Volume Maximization}

Following Stone, we feel that it is appropriate to first select the black and white points of the standard gamut, and then optimize the color.  Not only does this help reduce the number of degrees of freedom in the optimization procedure from twelve to six, but it makes sense from a human-computer interaction perspective as well: the prevalence of black and white text in computer displays makes it especially important to achieve a sharp visual distinction between the two colors.   To avoid visible color shifts, it is also important that the black and white points have
the same chromaticity, as in Stone's procedure.
Thus, we define the following objective function for optimizing a choice of standard gamut:
\begin{itemize}
\item Select $K$ and $W$ to be two points in the intersection of the projector gamuts,
both having the same chromaticities, and maximizing the ratio of the luminosities
of $K$ and $W$.
\item Choose a standard gamut with maximum volume among all gamuts that have the given values of $K$ and $W$ and that are contained within all the projector gamuts.
\end{itemize}

Both of these choices are somewhat arbitrary, and as we will see could be replaced by other similar objectives without significantly changing our optimization algorithms.  For instance, other objectives than the ratio of luminosities could be used for selecting $K$ and $W$, or they could be set independently of each other to minimize and maximize their luminosity values.  Compared with Stone's approach, though, we have the following important differences. First, the initial optimization of $K$ and $W$ is not altered by a final scaling step; instead, we set $K$ and $W$ once and for all and then optimize the remaining parameters subject to those constrained values.  Secondly, we have an explicit objective function that we seek to optimize (the volume of the gamut), stated independently of any algorithm for performing that optimization.

\subsection{Setting Black and White}
\label{sec:bw}

We now describe how to find values of $K$ and $W$ implementing the first part of our objective function for the optimal standard gamut.

Let $P$ denote the polyhedron, in three-dimensional device-independent color space, formed by intersecting the parallelepipeds formed by the color gamuts of the projectors in our tiled display.
If we intersect $P$ with a line $\ell_c$ through the origin, representing all points with a fixed chromaticity $c$, then this line will intersect $P$ at two points:
$\lambda^-(c)$ is the closest point to the origin in $P\cap\ell_c$, and has the minimum luminosity of any point in $P$ with color $c$;
$\lambda^+(c)$ is the farthest point from the origin in $P\cap\ell_c$, and has the maximum luminosity of any point in $P$ with color $c$.
Our task is to choose $c$ in such a way as to maximize the ratio of the luminosities of $\lambda^+(c)$ and $\lambda^-(c)$.

\begin{figure}[t]
\centering
\includegraphics{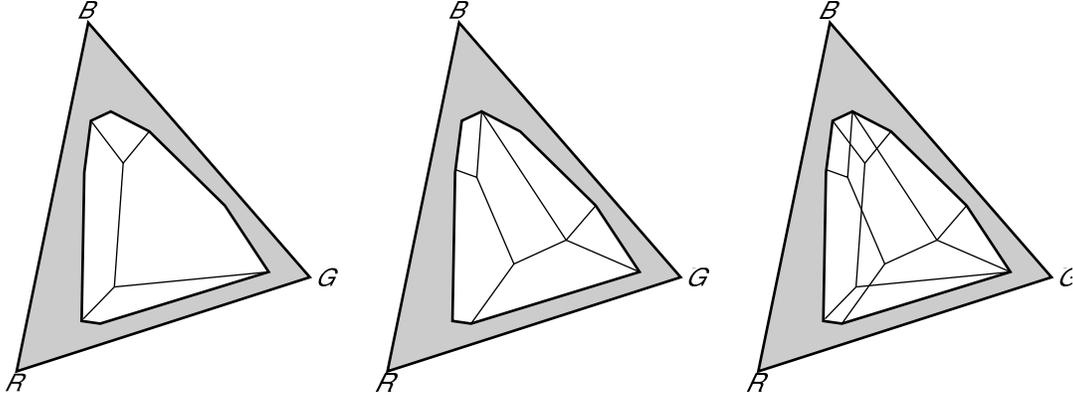}
\caption{Projections of the gamut intersection $P$ onto the chromaticity diagram of the color space.  Left: faces of $P$ nearest the origin, showing the face of $P$ determining $\lambda^-(c)$ for each $c$.
Center: faces of $P$ farthest from the origin, showing the face of $P$ determining $\lambda^+(c)$ for each $c$.
Right: overlaying the near and far diagrams.  The optimal $c$ must be a vertex of this overlaid diagram.}
\label{fig:overlaybw}
\end{figure}

To see how to do this, consider projecting the gamut intersection $P$ onto a plane of constant luminosity, forming a convex polygon $\bar P$ within the triangular chromaticity diagram of our color space.
Partition the facets of $P$ into two subsets $P^-$ (the facets visible from the origin, containing points $\lambda^-(c)$)
and $P^+$ (the facets on the far side of the origin from $P$, containing points $\lambda^+(c)$).
Then projecting $P^-$ onto the chromaticity diagram forms a subdivision of $\bar P$ into smaller convex polygons (Figure~\ref{fig:overlaybw}, left), and projecting $P^+$ onto the chromaticity diagram forms a different subdivision of $\bar P$ into convex polygons
 (Figure~\ref{fig:overlaybw}, center).  Overlaying the two projections produces a third, more complicated diagram $\hat P$ (Figure~\ref{fig:overlaybw}, right).
 
 \begin{lemma}
 The optimal value of $c$ is one of the vertices of diagram $\hat P$.
 \end{lemma}
 
 The proof idea is that within any of the higher dimensional faces of the diagram, the value we are trying to optimize is a fractional linear function that cannot have any local optima; we omit the details.  Thus, we are led to the following algorithm for finding $K$ and $W$, which implicitly enumerates all the vertices of this diagram. 
  
 \begin{enumerate}
 \item Construct a representation of the vertices, edges, and facets of polyhedron $P$ by intersecting the gamuts of the individual projectors.  Partition the boundary of $P$ into $P^-$ and $P^+$.
 \item For each vertex $v$ of $P$ in $P^-$, let $c_v$ be the chromaticity of the vertex;
 the vertex itself is $v=\lambda^-(c_v)$.   Construct the line through the origin with fixed chromaticity $c_v$, and intersect this line with the planes through each facet of $P^+$.
 Let $\lambda^+(c_v)$ be the nearest of these intersection points to the origin.
 \item Similarly, for each vertex $w=\lambda^+(c_w)$ in $P^+$, 
 construct the line with fixed chromaticity $c_w$, intersect this line with the planes through each facet of $P^-$, and let $\lambda^-(c_w)$ be the farthest of the resulting intersection points from the origin.
 \item For each two line segments $s\in P^-$ and $t\in P^+$,
 project these two segments into the chromaticity diagram.  If they intersect in a single point $c_{st}$, let $\lambda^-(c_{st})$ denote the point of chromaticity $c$ on $s$,
 and let $\lambda^+(c_{st})$ denote the point of chromaticity $c$ on $t$.
 \item Compare the luminosity ratios of all the pairs $\{\lambda^-(c),\lambda^+(c)\}$ constructed in steps 2--4, let $c^*$ denote the value of $c$ maximizing this ratio among these pairs,
 let $K=\lambda^-(c^*)$, and let $W=\lambda^+(c^*)$.
 \end{enumerate}
 
 \begin{lemma}
 The algorithm described above finds the optimal values of $K$ and $W$ in $O(n^2)$ time.
 \end{lemma}

We note that a similar diagram overlay technique can be used to find the optimal choice of $K$ and $W$ (constrained to have a common chromaticity) for many other objective functions than our choice of the luminosity ratio.
It seems likely that the time for these steps can be reduced using techniques for optimization in overlays of two planar diagrams~\cite{AgaShaTol-Algs-94}, however this would not reduce the complexity of our overall algorithm.

\subsection{Filling out the gamut}
\label{sec:volmax}

We now discuss the more difficult algorithmic problem of computing the maximum-volume gamut for a fixed choice of $K$ and $W$.  As discussed in Section~\ref{sec:color}, this gamut can be determined by a suitable set of four of its eight extremal colors.  Once $K$ and $W$ are chosen, we need only fix two more noncomplementary colors, say $R$ and $B$; all other extremal colors can be computed by formulas that are linear in the coordinates of $R$, $K$, $B$, and $W$.
However, $R$ and $B$ can be chosen independently among the points of the device-independent color space.
Thus, the set of possible gamuts that we must search forms a six-dimensional space in which $R$ and $B$ each contribute three coordinates.  We denote points in this space by pairs $(R,B)$ where each of $R$ and $B$ are themselves three-dimensional colors.

\begin{lemma}
\label{lem:6d-halfspaces}
In the six-dimensional space $(R,B)$, the set of gamuts that are contained in the intersection of all projector gamuts corresponds to a polytope $\Gamma$ formed by the intersection of $36n$ halfspaces, where $n$ denotes the number of projectors in the system.
\end{lemma}

\begin{proof}
If $P$ and $Q$ are any two convex polytopes, then $P\subset Q$ if and only if the vertices of $P$ all belong to $Q$.  Thus, a standard gamut determined by the four extremal colors $K$, $W$, $R$, and $B$ is contained within the intersection of the printer gamuts if and only if all eight extremal colors are contained within this intersection.  The two colors $K$ and $W$ are guaranteed by our contruction to already be within this intersection.  For any other color
$c\in\{R,G,B,C,M,Y\}$, any projector gamut $\Gamma_i$, and any facet $f$
among the six facets of $\Gamma_i$, the plane through $f$ must not separate $c$ from $\Gamma_i$.   Since each color $c$ can be expressed as a linear combination of the coordinates of $R$ and $B$ (and of the constant coordinates of $K$ and $W$), the constraint that $c$ and $\Gamma_i$ must be on the same side of $f$ can be expressed as a linear inequality
in the six-dimensional space $(R,B)$.
There are six choices of $c$, $n$ choices of $\Gamma_i$, and six choices of $f$ for each $\Gamma_i$, so the total number of these linear inequalities is $36n$.
\end{proof}

We next consider the nature of the objective function we are trying to optimize.
On the face of it, this is a cubic function:
the determinant of the $3\times 3$ matrix the rows of which are the three-dimensional vectors
$R-K$, $B-K$, and $G-K$.
However, we can simplify this considerably: by additivity of determinants, and the fact that
$G-K=(W-K)-(R-K)-(B-K)$,
we can separate the volume determinant into a sum of three determinants of simpler matrices
$\vol(R,B)=\det(R-K,B-K,W-K)-\det(R-K,B-K,R-K)-\det(R-K,B-K,B-K)=\det(R-K,B-K,W+K)$.
Since $W-K$ is already fixed, the function to maximize over $\Gamma$
is therefore seen to be a quadratic over the coordinates $(R,B)$.
Unfortunately, this quadratic is neither convex nor concave.  If it were convex, we could find the optimal standard gamut as a convex program, using generalized linear programming techniques, in time $O(n)$~\cite{MatShaWel-Algo-96}, while if it were concave, the optimal gamut would be a vertex of $\Gamma$, and could be found by enumerating all vertices of $\Gamma$ and computing the objective function for each.

However, the quadratic nature of $\vol(R,B)$ does help us somewhat:
suppose we want to determine the maximum of $\vol(R,B)$ within a compact convex set $S$.
Find the affine hull of $S$ (that is, the smallest linear subspace of the six-dimensional space $(R,B)$ that contains all of $S$), and restrict the function $\vol$ to this subspace.  If the restricted quadratic is positive definite or indefinite, it is unbounded in the linear subspace, and the maximum must occur on the boundary of $S$.  If the restricted quadratic is negative definite, it has a unique local maximum on the linear subspace.  If this maximum occurs within the interior of $S$, it is the maximum on $S$ as well as on the larger linear subspace.  Otherwise, the maximum again must occur on the boundary of $S$.  We can use the facts in the following algorithm
for finding $(R,B)$ maximizing $\vol(R,B)$:

\begin{enumerate}
\item Find the set of $36n$ 6-dimensional halfspaces,
described in Lemma~\ref{lem:6d-halfspaces},
such that any pair $(R,B)$ contained within the intersection of these halfspaces
determines a feasible standard gamut.
\item Construct the polytope $\Gamma$ formed by intersecting the halfspaces. The polytope's worst case complexity is $O(n^3)$ by McMullen's upper bound theorem~\cite{McM-M-70,McMShe-71}, and worst-case optimal halfspace intersection algorithms (by projective duality, equivalent to convex hull construction) with time matching this complexity are known~\cite{Cha-DCG-93,Sei-MS-81}.
\item Subdivide the boundary of $\Gamma$ into simplices (triangles, tetrahedra, and their higher dimensional analogues).  This can be done by recursively subdividing all lower dimensional boundary faces, and then within each facet connecting the bottommost vertex to each simplex in the triangulation of the facet's boundary.
This {\em pulling triangulation} again has $O(n^3)$ complexity by the upper bound theorem.
\item Within each $k$-dimensional simplex $\Delta$ of the triangulation, restrict the function $\vol(R,B)$ to the affine hull of $\Delta$ by substituting linear equations for $6-k$ of the coordinates of $(R,B)$.  Test whether the resulting restricted quadratic function is negative definite.  If so, and the maximum of the function is within $\Delta$, then let $(R_\Delta,B_\Delta)$ denote the point achieving this maximum.
\item Compute $\vol(R_\Delta,B_\Delta)$ for all points $(R_\Delta,B_\Delta)$ constructed in the previous step, and let $(R,B)$ denote the pair achieving the maximum volume.
\end{enumerate}

Other objective functions could be substituted, as long as we can find optimal values of the function within six-dimensional simplices, without increasing the overall complexity of the algorithm.

\subsection{Summary of Algorithm}

To summarize, our volume maximization algorithm consists of finding
$K$ and $W$ by overlaying two chromaticity diagrams as described
in Section~\ref{sec:bw},
finding $R$ and $B$ by searching for maxima of $\vol(R,B)$ within each simplex of
a triangulation of the six-dimensional polytope $\Gamma$
as described in Section~\ref{sec:volmax},
and using $K$, $W$, $R$, and $B$ to determine the standard gamut.

\begin{theorem}
\label{thm:exact-vol}
The algorithm described above finds a standard gamut optimizing our objective function in time $O(n^3)$, where $n$ denotes the number of projector gamuts in the input.
\end{theorem}

We note that further speedups are possible if we allow for approximate results:
a $(1-\epsilon)$-approximation to the volume-maximizing gamut for given corners $K$ and $W$ can be found in time $O(n\log n + \epsilon^{-3})$
by replacing the intersection of the projector gamuts with an approximate intersection
having $O(1/\epsilon)$ vertices.  This approximate intersection can be found by using an affine transformation to ensure that the true intersection is sufficiently fat, then applying
an algorithm of Dudley~\cite{AgaHarSha-JACM-97,Dud-JAT-74} to replace the intersection by a polytope with fewer vertices; we omit the details.

\section{Quasiconvex Programming}

Our discussion above centers on the optimization of objective functions involving volume of the color gamut.  However, our algorithms could easily be adapted to other objective functions, and it is not clear that volume is the correct choice.  We see two potential problems with our volume-based approach. First, the running time, while cubic, has a high constant factor due to the factor of $36$ in Lemma~\ref{lem:6d-halfspaces}; this may make our running times impractically slow.
This speed issue could be addressed from the point of view of approximation algorithms, as in the previous section, but it might be more appropriate to find a faster exact algorithm for an easier-to-optimize quality measure.  Second, volume maximization neglects another issue of color gamut optimization: how do we match the corners of the chosen standard gamut to the usual additive primary colors $R$, $G$, and $B$?  Any permutation of these corners would produce a geometrically equivalent parallelepiped but only one of these six permutations is likely to be appropriate for use as our standard gamut.  In practice, we expect the choice of corner labeling to be obvious, but this ambiguity indicates a possible weakness of our volume-based approach.  In this section, we present a more generic approach based on quasiconvex programming that would have the advantages of faster running time and automatic gamut corner labeling; however more colorimetric expertise would be needed to make specific design choices among those offered by this approach.

{\em Quasiconvex programming} is a general framework for LP-type computational geometry optimization problems, formulated by Amenta et al.~\cite{AmeBerEpp-Algs-99} in the context of mesh smoothing procedures, and later used by the authors for several applications in information visualization and mesh generation~\cite{BerEpp-WADS-01}.
Define a {\em nested convex family} to be a function $\kappa(t)$ mapping real values $t$ to convex sets in some Euclidean space $\E^d$, such that for any $t<t'$, $\kappa(t)\subseteq\kappa(t')$, and such that for all $t$, $\kappa(t)=\bigcap_{t'>t}\kappa(t')$.  If $\kappa$ is a nested convex family, we can define a quasiconvex function
$f_\kappa(x) = \inf\,\{\,t \mathrel{|} x \in \kappa(t)\,\}$
mapping $\E^d$ back to the real numbers; conversely every quasiconvex function corresponds to a nested convex family.
An instance of a {\em quasiconvex program} consists of a finite set $K=\{\kappa_0,\kappa_1,\ldots\}$ of nested convex families (or equivalently quasiconvex functions).  The desired output is the pair
$$
\inf\Big\{\,(t,x) \mathrel{\big|}
x\in \mathop{\textstyle\bigcap}\limits_{\kappa_i
\in K}\kappa_i(t)
\Big\}
$$
where the infimum is taken in the lexicographic ordering,
first by $t$ and then by the coordinates of~$x$.
Less formally, we seek the point $x$ minimizing the maximum value of
$f_{\kappa_i}(x)$.

As Amenta et al.~\cite{AmeBerEpp-Algs-99} showed, when $d$ is a fixed constant, any quasiconvex program consisting of $n$ nested convex families can be solved by generalized linear programming techniques, assuming the existence of primitives  for evaluating the input quasiconvex functions as well as for solving subproblems with a constant number of inputs.  The time for these techniques is $O(En + S\log n)$ where $E$ is the time for an evaluation and $S$ is the time for a subproblem solution.  Alternatively, Amenta et al.{} showed that simple hill-climbing techniques for quasiconvex programs are guaranteed to converge to the globally optimal value.

To formulate the color gamut optimization problem as a quasiconvex program,
recall that additive gamuts can be parametrized by a four-tuple of colors, say $(K,R,G,B)$,
and that the other four colors $C, M, Y, W$ can be expressed as linear functions of these four colors.
We need functions $q_c(x)$ for each gamut corner $c\in\{K,R,G,B,C,M,Y,W\}$,
measuring how well a point $x$ in device independent color space matches some desired ideal location for $c$.  These functions $q_c$ could be Euclidean distances from ideal locations,
or linear objective functions measuring distance along ideal directions ("blackest black", "reddest red", and so forth).  It may be appropriate to assign greater weight to the locations of $K$ and $W$ than to the other corners.
Lower values of $q_c$ will indicate a better fit to the ideal location, and higher values will indicate a worse fit.
We will use these functions to measure the quality of a gamut $\Gamma$,
in the 12-dimensional space of color gamuts parametrized by color 4-tuples $(K,R,G,B)$,
where the quality $Q_c(\Gamma)$ is measured by finding the point labeled $c$ in gamut $\Gamma$ and applying $q_c$ to that point.

\begin{lemma}
$Q_c$ is quasiconvex if and only if $q_c$ is quasiconvex.
\end{lemma}

Thus, in order to apply quasiconvex programming, all we need is for each of the functions $q_c$ to be quasiconvex.

\begin{theorem}
Suppose that for each  $c\in\{K,R,G,B,C,M,Y,W\}$ we are given a quasiconvex function $q_c$ on device-independent color space, and define $Q_c$ as above.  Then the problem of finding the gamut $\Gamma$ minimizing $\max_c Q_c(\Gamma)$ can be solved in linear time by quasiconvex programming.
\end{theorem}

It is natural to hope that colorimetric principles will lead to a good choice of the exact form of the functions $q_c$; we leave the details of this choice for further work.
We note that the formulation above does not include any constraint that $W$ and $K$ have the same chromaticities; it would be of interest to incorporate such a constraint into our quasiconvex programming framework, or to determine whether the color shifting it prevents would actually be a problem in the unconstrained quasiconvex program.

\section{Conclusions and Further Work}

We have formulated problems of color gamut intersection as that of optimizing an objective function on parallelepipeds within a convex polyhedron, and found efficient algorithms for this optimization task for several colorimetrically motivated definitions of the objective function.
The algorithms described in this paper have not yet been implemented, but we believe they should be simple and efficient enough to be suitable for typical problem sizes; we hope to continue this work by implementing some of our algorithms and comparing their results with those of Stone's procedure.

More generally, color spaces seem likely to provide a fruitful source of problems in three-dimensional computational geometry.  Other problems that have been studied in this area include color gamut mapping~\cite{HarSch-CIC-97}, color quantization~\cite{InaKatIma-SCG-94,VelGomSob-SCG-98}, and optimized printing using restricted numbers of ink colors~\cite{PowWesSto-SIGGRAPH-96}
%% , and selection of contrasting but non-clashing highlight colors~\cite{},
%% I looked up Rosenholtz's pubs and didn't find anything that looked
%% right for this - DE 12/2/02
but it seems likely that much more work remains to be done in the computational geometry of color.

\section*{Acknowledgements}

Work of Eppstein was done in part while visiting Xerox
PARC and supported in part by NSF grant CCR-9912338.
We thank Maureen Stone for helping us understand the mysterious world of color.

\raggedright
\bibliographystyle{abuser}
\bibliography{gamut}
 \end{document}